\documentclass[9pt,twocolumn,twoside]{opti_preprint}
\setboolean{shortarticle}{false}

\usepackage{color}
\usepackage{amsmath} 
\newcommand\rr{\mathbf{r}}

\title{Super-resolution photoacoustic imaging via flow induced absorption fluctuations}

\author[1,$\dagger$]{Thomas Chaigne}
\author[2,3,$\dagger$]{Bastien Arnal}
\author[2,3]{Sergey Vilov}
\author[2,3]{Emmanuel Bossy}
\author [4,*] {Ori Katz}

\affil[1]{Bioimaging and Neurophotonics Lab, NeuroCure Cluster of Excellence, Charité Berlin, Humboldt University, Charitéplatz 1, 10117 Berlin, Germany}
\affil[2]{Univ. Grenoble Alpes, LIPHY, F-38000 Grenoble, France}
\affil[3]{CNRS, LIPHY, F-38000 Grenoble, France}
\affil[4]{Department of Applied Physics, Hebrew University of Jerusalem, Israel}

\affil[$\dagger$]{These authors contributed equally to this work.}
\affil[*]{Corresponding author: orik@mail.huji.ac.il}

\begin{abstract}
Abstract - In deep tissue photoacoustic imaging the spatial resolution is inherently limited by the acoustic wavelength. We present an approach for surpassing the acoustic diffraction limit by exploiting temporal fluctuations in the sample absorption distribution, such as those induced by flowing particles. In addition to enhanced resolution, our approach inherently provides background reduction, and can be implemented with any conventional photoacoustic imaging system. The considerable resolution increase is made possible by adapting notions from super-resolution optical fluctuations imaging (SOFI) developed for blinking fluorescent molecules, to flowing acoustic emitters.
By generalizing SOFI mathematical analysis to complex valued signals, we demonstrate super-resolved photoacoustic images that are free from oscillations caused by band-limited detection.
The presented technique holds potential for contrast-agent free micro-vessels imaging, as red blood cells provide a strong endogenous source of naturally fluctuating absorption.

\end{abstract}

\begin{document}

\maketitle
\ifthenelse{\boolean{shortarticle}}{\abscontent}{}

\section{Introduction}
Optical microscopy is an invaluable tool in biomedical investigations and clinical diagnostics. However, its use is limited to depths of not more than a few hundreds micrometers inside tissue due to light scattering. This fundamental limitation originates from the irremediable optical wavefront distortions that are induced by light scattering, preventing diffraction-limited optical focusing at larger depths. While deeper investigations are possible using either tomographic techniques based on optical diffusion or by combining non-optical modalities such as ultrasound, these deeper-penetrating techniques do not allow optical diffraction-limited resolution, and generally do not permit microscopic studies of e.g. cellular structures at depths of more than a millimeter \cite{ntziachristos_going_2010}. One of the leading approaches for deep tissue ultrasound-aided optical imaging is photoacoustic (PA) imaging  \cite{wang_photoacoustic_2012, beard_biomedical_2011}. Photoacoustic imaging relies on the generation of ultrasonic waves by light absorption in a target structure under pulsed optical illumination. The ultrasonic waves that are produced in the absorbing parts of the target via thermo-elastic stress generation, propagate in soft tissue almost without scattering, and are detected by an external ultrasonic detector. PA is thus a noninvasive imaging technique which provides images of optical absorption at large depth with a spatial resolution limited by acoustic diffraction. Ultimately, the ultrasound resolution in soft tissue is limited by the acoustic wavelength used, which is limited by the attenuation of high frequency ultrasonic waves. As a result, the depth-to-resolution ratio of deep-tissue PA imaging is approximately 200 in practice \cite{wang_photoacoustic_2012}. For example, at a depth of 5 mm a resolution of around 25 $\mu$m at best may be expected, two orders of magnitude above the optical diffraction limit \cite{omar_ultrawideband_2014}. Such a resolution is insufficient for the important tasks of deep-tissue cellular imaging and for the identification of angiogenic microvessels in the microvasculature \cite{mcdonald2003imaging}.

Recently, it was demonstrated that the conventional acoustic diffraction-limit can be surpassed by exploiting temporal fluctuations in PA signals that originates from dynamic optical speckle illumination \cite{chaigne_super-resolution_2016}. Inspired by super-resolution optical fluctuation imaging (SOFI) \cite{dertinger_fast_2009}, where blinking fluorescent molecules are used to induce temporal fluctuations of fluorescence, multiple random speckle illuminations were exploited as a source of PA fluctuations \cite{chaigne_super-resolution_2016}. The acoustic diffraction limit was surpassed by statistical analysis of a set of PA images obtained under such set of random, unknown optical speckle patterns.
While impressive results based on this approach have been reported in several recent works using tissue phantoms \cite{hojman_photoacoustic_2017,murray_super-resolution_2017}, a major practical hurdle prevents the use of optical speckle-based techniques for deep tissue PA imaging. This limit is set by the low contrast of the PA fluctuations amplitude, resulting from the large difference between the small optical speckle grain size to the acoustic wavelength used at depths larger than a millimeter. At such depths the optical speckle grain dimensions are of the optical wavelength scale (of the order of one micrometer), orders of magnitude smaller than the ultrasound wavelength (typically 100 µm for a typical frequency of about 15 MHz). This mismatch in dimensions leads to a small amplitude of fluctuations with respect to the mean value in each image pixel, since the PA signal at each image pixel is the sum of a large number of uncorrelated fluctuating speckle grains enclosed within the large acoustic point spread function (PSF) \cite{goodman2007speckle,gateau_improving_2013}. Resolving the small fluctuations over the large mean signal is challenging under common signal to noise (SNR) conditions \cite{gateau_improving_2013}. 

In this work, we present a novel PA technique that provides a resolution beyond the acoustic diffraction limit without relying on any structured or coherent illumination. By exploiting fluctuations in the sample itself rather than external illumination-induced fluctuations, it allows to overcome many of the difficulties and limitations of the state of the art approaches. Specifically, we demonstrate that fluctuations in the PA signals that are induced by the \textit{motion} of absorbers leads to super-resolved PA images. In biological tissue, naturally moving endogenous absorbers may consist of red blood cells flowing through the microvasculature, an important imaging target in e.g. the evaluation of drugs that affect angiogenesis \cite{mcdonald2003imaging}.
In this context of flowing absorbers, flowing \textit{exogenous} contrast agents have been very recently exploited for super-resolved imaging in medical ultrasound, either by localization of sparse distribution of absorbers \cite{errico2015ultrafast}, or by adapting a SOFI-based statistical analysis \cite{bar2017fast}. In photoacoustics, sparsely distributed flowing particles have recently been proposed \cite{dean2017dynamic} as a replacement for speckle illumination \cite{gateau_improving_2013} for improving visibility and contrast in limited-view PA tomography. Here, by employing a SOFI-based statistical analysis framework, we demonstrate that non-sparse flowing absorbers can be exploited for super-resolved background-reduced PA imaging. Furthermore, we extend the SOFI mathematical framework to allow processing of complex-valued acoustic signals, which is an important crucial step in order to discard the PA imaging artifacts that are caused by the typical axial oscillations of the PA band-limited PSF. By exploiting optical absorbers, our simple super-resolution approach possesses the advantage of optical and potentially endogenous contrast, such as the optical absorption of red blood cells, which is not available in ultrasound only imaging. In this context it is important to note that the SOFI-based processing enables super-resolved imaging even with non-sparse distributions of sources \cite{dertinger_fast_2009}. Here we demonstrate these advantages using conventional PA images of flowing absorber, which can be acquired with commercially available systems.

\section{Principle and numerical simulations}
\begin{figure}[h!]
\centerline{\includegraphics[width=0.95\columnwidth]{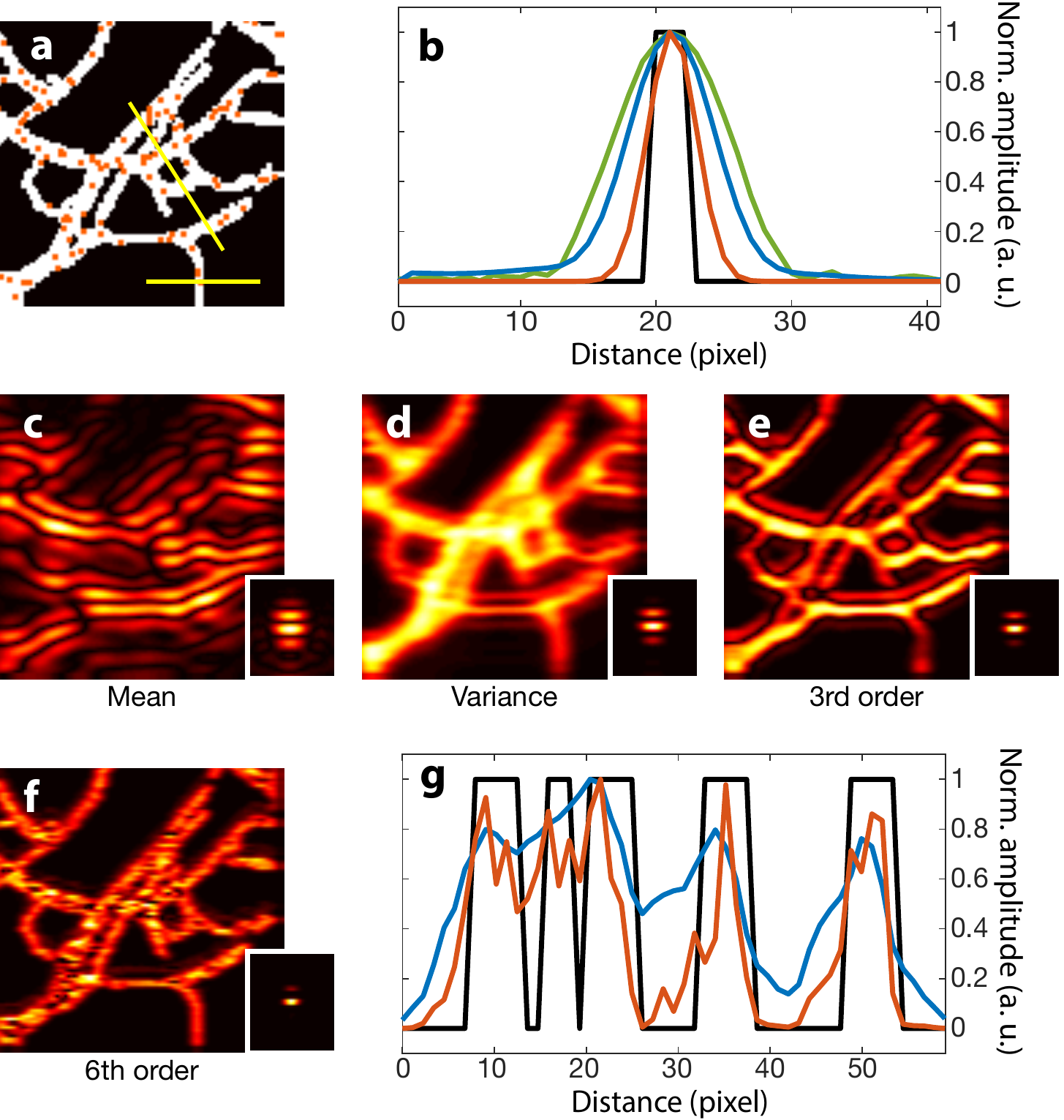}}
\caption{Numerical results with vasculature-like phantom. (a) Vessel-like structure (white), with small circulating absorbers (orange). The yellow lines indicate the directions used for the cross-sections in (b) and (g). (b) Cross-sections along the horizontal yellow line in (a): black: target structure, blue: variance image, orange: $6^{th}$-order cumulant. Green: horizontal cross-section of the main lobe of the PSF (inset in (c)). (c) Absolute value of mean PA image, averaged over all acquired images of flowing absorbers. Inset: absolute value of the PSF, $|h(\rr)|$. (d) Variance image. Inset: $|h^2(\rr)|$. (e) Absolute value of $3^{rd}$-order cumulant. Inset: $|h^3(\rr)|$. (f) Absolute value of $6^{th}$-order cumulant. Inset: $|h^6(\rr)|$. (g) Cross-sections along the oblique yellow line in (a): black: target vessel-like structure, blue: variance image, orange: $6^{th}$-order cumulant.}
\label{bvz}
\end{figure}

To explain the source of increased resolution in our approach, we consider the following model of PA imaging: in its simplest form, PA imaging with short pulse illumination may be described as a convolution of the spatial distribution of absorbed optical energy $E_{\mathrm{abs}}(\mathbf{r})=I(\mathbf{r})\times\alpha(\mathbf{r})$ (with $I(\mathbf{r})$ the illumination distribution and $\alpha(\mathbf{r}$) the distribution of optical absorption) with the PSF $h(\mathbf{r})$ of the imaging system. For simplicity, we consider a constant and uniform illumination $I(\mathbf{r})=I_0$ for each laser shot, and a temporally varying random distribution of absorption, given for each of the $k=1..N$ laser shots by 
\begin{equation}
\label{eq:absdistribution}
\alpha_{k}(\mathbf{r})=f(\mathbf{r})\times g_k(\mathbf{r})
\end{equation}In the above model, $g_k(\mathbf{r})$ describes the random distribution of moving optical absorbers at the time of the $k^{th}$ laser shot and $f(\mathbf{r})$ the geometry of the structure through which optical absorbers are moving, i.e. the target structure for imaging. For PA imaging of blood vessels, $f(\mathbf{r})$ would represent the vascular structure (Fig. \ref{bvz}(a), white) and $g_k(\rr)$ the random distribution of red blood cells in the vascular structure at laser shot $k$ (Fig. \ref{bvz}(a), orange).
Thus, for each laser shot $k$ the PA image $PA_k(\rr)$ may be described by the following equation:
\begin{equation}
\label{eq:PAimage}
PA_k(\rr)=\left[I_0 \times f(\rr)\times g_k(\rr) \right]\ast h(\mathbf{r})
\end{equation}
Our approach consists of estimating the target structure $f(\rr)$ beyond the resolution limit of the acoustic PSF $h(\rr)$, by exploiting the temporal fluctuations in the unknown distribution of absorbers $g_k(\rr)$ flowing through the structure at the $k^{th}$ laser shot. 
Following the original SOFI approach \cite{dertinger_fast_2009}, an $n^{th}$-order statistical fluctuation analysis is performed on the acquired image set to obtain a $\sqrt{n}$ resolution increase without deconvolution. This is accomplished by calculating the $n^{th}$-order auto-cumulant $C_n(\rr)$  for each pixel in the PA image set. $C_n(\rr)$ is calculated recursively using the previous order cumulants and the $k=1..n^{th}$-order moments $\mu_k(\rr)=\dfrac{1}{N}\sum_{k=1}^{N}\left[PA_k(\rr)\right]^m$ \cite{dertinger_fast_2009}:
\begin{equation}
C_n(\rr)=\mu_n(\rr)-\sum_{k=1}^{n-1} \left ( \begin{matrix}n-1  \\ k-1 \end{matrix} \right)C_k(\rr)\mu_{n-k}(\rr)
\end{equation}where  $C_1(\rr)=\mu_1(\rr)$  is the mean image, and $C_2(\rr)$  provides the variance image. 
Following Dertinger et al. \cite{dertinger_fast_2009}, the $n^{th}$-order cumulant provides images that involve a convolution only with the $n^{th}$ power of the PSF rather than the PSF itself, and thus provide a $\sqrt{n}$ resolution increase for a Gaussian PSF, without deconvolution. Deconvolution could further increase the resolution \cite{dertinger_fast_2009,chaigne_super-resolution_2016}.

As a first study of the expected resolution increase, we performed numerical simulations of a dynamic random absorption distribution in a complex microvasculature structure. Fig. \ref{bvz}(a) shows a single distribution of the absorbers (in orange) ($g_k(\mathbf{r})$) within the vessel-like target (in white) ($f(\rr)$). For each random absorbers-distribution, a PA image $PA_k(\rr)$ was simulated by convolving $f(\rr)\times g_k(\rr)$ with a typical, experimentally measured PA imaging PSF of a linear ultrasound array (Fig. \ref{bvz}(c), inset), following Eq.\ref{eq:PAimage}. Fig. \ref{bvz}(c) shows the mean image, which is an estimate of $h(\rr)\ast f(\rr)$, and provides the conventional PA image. As can be appreciated the corresponding resolution is too low to resolve the vessel-like structure, as is the situation in many practical microvessel imaging tasks \cite{mcdonald2003imaging}. Fig. \ref{bvz}(d), (e) and (f) respectively show the $2^{nd}$-, $3^{rd}$- and $6^{th}$-order cumulant images calculated from a total of $N=90000$ simulated PA images of optical absorbers randomly distributed inside the vessel structure. The inset in each image illustrates the corresponding $n^{th}$-power of the PSF $h(\rr)$. These results clearly illustrate how high-order PA fluctuation analysis can provide a significant resolution increase, with the resolution increasing with the analyzed order $n$, as is illustrated in the one-dimensional cross-section through the images, shown in Fig. \ref{bvz}(b) and (g). While it is impossible to resolve the individual vessels even in the variance image in Fig. \ref{bvz}(g), the $6^{th}$-order cumulant allows resolving the individual vessels. As a quantitative measure, we have computed the full width at half maximum (FWHM) of the images of an isolated thin vessel (Fig. \ref{bvz}(b)) for the different analyzed cumulant orders, and compared it to the transverse FWHM of the PSF (inset in Fig. \ref{bvz}(c)). The resulting FWHM is 9.8 pixels for the PSF, 7.3 pixels for the variance (blue trace), and 4.2 pixels for the $6^{th}$-order cumulant (orange trace), in line with the theoretically expected $\sqrt[]{n}$ resolution increase \cite{dertinger_fast_2009}.

\section{Experimental results}
\begin{figure}[h!]
\centerline{\includegraphics[width=0.75\columnwidth]{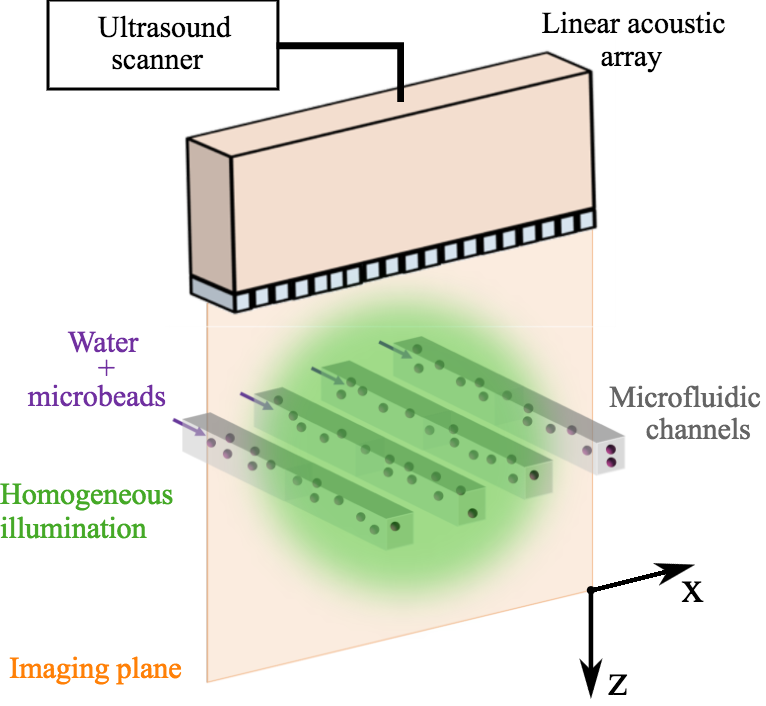}}
\caption{Schematic of the experimental setup: parallel microfluidic channels flown with a water suspension of absorbing beads (violet). The channels are positioned perpendicularly to the image plane (orange) of a linear ultrasound array. A pulsed laser beam (green) uniformly illuminates the imaging plane. For each nanosecond laser shot a PA image is acquired, where the flowing particles appear as a frozen random distribution of absorbers in the microfluidic channels, providing temporal PA fluctuations from one shot to the next.}
\label{setup}
\end{figure}

\begin{figure}[h!]
\centerline{\includegraphics[width=1.08\columnwidth]{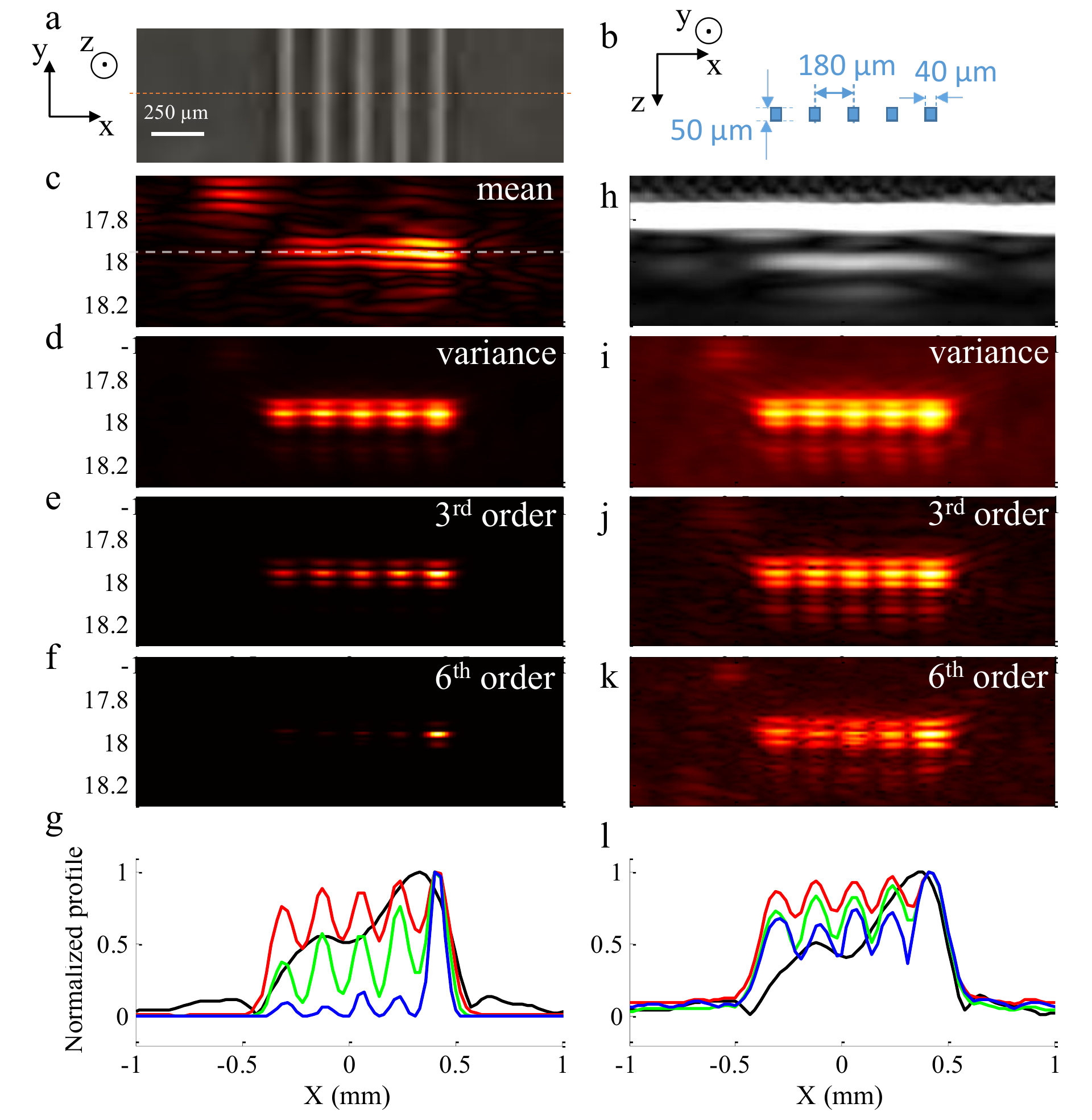}}
\caption{Experimental results. (a) Top-view photograph of the microfluidic circuit, showing the 5 parallel channels along the (x-z) imaging plane (dashed orange line). (b) Schematic of the channels with the relevant dimensions. (c) Mean PA image, representing the result of conventional PA imaging. The white dashed line indicates the direction of the cross-sections in (g) and (l). (d) Variance image ($2^{nd}$-order cumulant). (e) Absolute value of the $3^{rd}$-order cumulant. f) Absolute value of the  $6^{th}$-order cumulant. g) One-dimensional profiles across the channels for the mean (black), variance (red), $3^{rd}$-order (green) and $6^{th}$-order (blue) cumulant images. (h) Pulse-echo ultrasound image of the 5 channels filled with air, illustrating the inability to resolve the structure with conventional imaging. (i-k) $n^{th}$-root of the $n^{th}$-order cumulants for $n=2$ (i), $n=3$ (j) and $n=6$ (k), computed to provide images with comparable units. (l) One-dimensional profiles across the channels for the $n^{th}$-root of the $n^{th}$-order cumulant images, for $n=1$ (black), $n=2$ (red), $n=3$ (green) and $n=6$ (blue).}
\label{ExpResReal}
\end{figure}

To experimentally validate our approach we have performed a proof-of-principle experiment whose results are presented in Fig. \ref{ExpResReal}. A schematic of the experimental setup is shown in Fig. \ref{setup}. A Capacitive Micromachined Ultrasonic Transducer (CMUT) array (L22-8v, Verasonics, Kirkland, Washington, USA) connected to a multi-channel acquisition electronics (High Frequency Vantage 256, Verasonics, Kirkland, Washington, USA) was used to produce two-dimensional PA images of channels crossing the imaging plane.  PA signals with center frequency around 15 MHz were acquired simultaneously on 128 elements of the array (pitch 100 $\mu$m, elevation focus 15 mm) for each laser shot. The sample was illuminated with 5 ns laser pulses at 20 Hz repetition rate ($\lambda$ = 532 nm, fluence = 3.0 mJ/cm$^2$) from a frequency-doubled Nd:YAG laser (Spitlight DPSS 250, Innolas Laser GmbH, Krailling, Germany). Channels of a microfluidic circuit filled with a flowing suspension of absorbing particles were used as a model of red blood cells flowing inside vessels. The circuit was built in PDMS (polydimethylsiloxane) with standard soft-lithography manufacturing technology ~\cite{tang2010basic_bis}. The circuit plane was located at the (X-Y)-plane, 18 mm below the ultrasound transducer. 
The imaged part of the circuit sample consisted of 5 identical parallel channels, passing perpendicularly through the imaging plane (only 4 are represented on Fig. \ref{setup}). 
The 5 parallel channels were identical circuit sections arranged in series along the main circuit. To insure temporal decorrelation of the random stream of particle between the 5 sections, as required by the SOFI analysis, intermediate circuit elements with branching were used to shuffle the flow of absorbing particles. The center-to-center distance between neighbouring channels was 180 $\mu$m, and each channel was 40-$\mu$m wide (along the X direction) and 50-$\mu$m high (along the Z direction), (Fig. \ref{ExpResReal}(b)). The circuit was filled with pure water loaded with a suspension of monodisperse spherical 10 $\mu$m diameter absorbing particles  (microparticles GmbH, Berlin, Germany). The particle concentration was approximately 9100 particles per mm$^3$, corresponding to about 13 particles per channel on average in each PA image, given by the 720 {$\mu$m} thick imaging plane defined by the full elevational width at half maximum of the PA PSF.

For each nanosecond laser pulse, corresponding to one "frozen" random distribution of the particles flowing through the channels, the PA radio-frequency (RF) signals were acquired. The laser jitter and pulse energy variations were compensated using a fixed reference absorber located outside the region of interest. Images were reconstructed from the raw recorded acoustic signals, using a conventional delay-and-sum algorithm, assuming a homogeneous sound speed of $1500 m/s$. Electromagnetic noise from the laser results in  parasitic signals appearing mostly synchronously on all the elements. To enhance the signal-to-noise ratio, the mean RF value across the elements was subtracted on the RF data, for each time and frame before beamforming. The thin (200 $\mu$m-thick) PDMS layer overlaying the channels was neglected in the reconstruction algorithm. 
To enhance SNR and reduce PA background features, clutter filtering based on spatio-temporal singular value decomposition \cite{demene_spatiotemporal_2015} was performed on multiple series of 500 images ensembles keeping singular values from 2 to 8.
Statistical analysis was performed on 4000 beamformed images, and its results are presented in Fig. \ref{ExpResReal}(c-g,i-l). At the probe center frequency of 15 MHz, the corresponding lateral resolution (along the X direction) was about 140 $\mu$m. As result of the thin PDMS layer ultrasound absorption, the lateral resolution (along x direction) was degraded to approximately 200 µm, too coarse to resolve the individual channels (Fig. \ref{ExpResReal}(c),(h)). 

The inability to resolve the target structures with conventional ultrasound imaging is illustrated in Fig. \ref{ExpResReal}(h), where a conventional pulse-echo image of the 5 channels filled with water obtained with a single plane-wave emission is shown. The unresolved channels appear as a continuous line at z = 17.96 mm (the saturated line at z = 17.8 mm corresponds to the strong  reflection caused by the PDMS/water interface). This pulse-echo image corresponds to the optimal situation in term of SNR 
and has the exact same resolution as each of the conventional PA images.
The mean clutter filtered PA image Fig. \ref{ExpResReal}(c) shows features with strong PA signals, which do not correspond to channels: visual inspection of the circuit after the experiment suggested that various optical absorbers (possibly dust or clusters of absorbing beads) were fixed  to the circuit and produced the strong PA background features that were reduced by clutter filtering but not totally suppressed.
The $n^{th}$-order cumulant images for n=2, 3 and 6 are shown in Fig. \ref{ExpResReal}(d), (e) and (f), with the corresponding one-dimensional profiles across the line of channels shown in Fig. \ref{ExpResReal}(g), and validate experimentally the expected resolution increase. In addition to the resolution increase, these results also demonstrate that high-order fluctuation PA imaging is a powerful approach in suppressing background artifacts from stationary distributions of absorbers: the strong background features observed on the mean image (Fig. \ref{ExpResReal}(c)), which preclude the observation of the channels, are suppressed in the cumulants images.
In the context of biological tissue imaging, this may provide an effective approach to selectively image blood flow with super-resolution. Since the $n^{th}$-order cumulant are non-linearly related to the absorption profile \cite{dertinger_fast_2009}, Fig. \ref{ExpResReal}(i-k) presents the $n^{th}$-root of the $n^{th}$-order cumulant image. The $n^{th}$-root compensate for the nonlinear over-weighting of the high order cumulants for the objects with the strongest signals, and a quantity linearly related to the absorption of the object is recovered. It is important to note that even when the $n^{th}$-root is presented, the channels are increasingly better-resolved at higher orders (Fig. \ref{ExpResReal}(l)). This enhanced resolution with increasing order depends on the  $n^{th}$-order statistics of the flowing particles and can be more or less effective in practice depending on the actual statistics of the flowing particles. 

Inspecting the PA images presented in Fig. \ref{ExpResReal} (i-k), one can clearly notice the substantial artefacts in the form of anisotropic axial oscillations in the PA signal. These oscillations results from the bandwidth limited bipolar impulse response of the ultrasound signals, and thus the acoustic PSF. 
In conventional PA and ultrasonic imaging this typical oscillating artefacts are avoided by displaying the envelope of the complex analytical  signal or magnitude of complex-valued images. Such complex-value processing is a crucial step in avoiding any misinterpretation of PSF oscillations as real structures. 
To retrieve a similar envelope- or magnitude-related information in the presented SOFI-based fluctuation analysis, one can not simply perform a Hilbert transform on the cumulant images, which are nonlinearly related to the raw signals. For example cumulants of even orders are by definition non-negative. Thus we develop below an analytical signal processing that is based on a generalization of the real value moments and cumulants definitions for complex random variables \cite{eriksson_essential_2010}. Complex-valued PA images were obtained by beamforming the complex analytical temporal signals (obtained with a Hilbert transform of the raw real-valued RF signals). We defined $n^{th}$ order complex moments as: $m_{p,q}(\rr)=\dfrac{1}{N}\sum_{k=1}^{N}\left[PA_k(\rr)]^p.[PA_k(\rr)^*\right]^q$ where $q = n-p$.
The complex cumulants of order $n=p+q$, $K_{p,q}$, are linked to the complex moments, $m_{p,q}$ by  the following formula derived from \cite{eriksson_essential_2010}. For $q>0$:
\begin{flalign}
\label{eqrecurs}
K_{p,q}(\rr) = m_{p,q}(\rr) - \sum_{u=1}^{p}   \sum_{v=1}^{q-1} \left ( \begin{matrix}p  \\ u \end{matrix} \right) \left ( \begin{matrix}q-1  \\ v \end{matrix} \right)K_{p-u,q-v}(\rr)m_{u,v}(\rr) \\\nonumber
- \sum_{v=1}^{q-1} \left ( \begin{matrix}q-1  \\ v \end{matrix} \right) K_{p,q-v}(\rr)m_{0,v}(\rr) 
- \sum_{u=1}^{p} \left ( \begin{matrix}p  \\ u \end{matrix} \right) K_{p-u,q}(\rr)m_{u,0}(\rr) \\\nonumber
\end{flalign}
If $q=0$, the following formula applies:
\begin{equation}
K_{p,0}(\rr)=m_{p,0}(\rr)-\sum_{u=1}^{p-1} \left ( \begin{matrix}p-1  \\ u-1 \end{matrix} \right) K_{u,0}(\rr)m_{p-u,0}(\rr)
\end{equation}where $p+q$ is the order of the cumulant. There are $n+1$ cumulants for a given order and it can be noted that $K_{p,q} = K_{q,p}*$.
\begin{figure}[h!]
\centerline{\includegraphics[width=1.0\columnwidth]{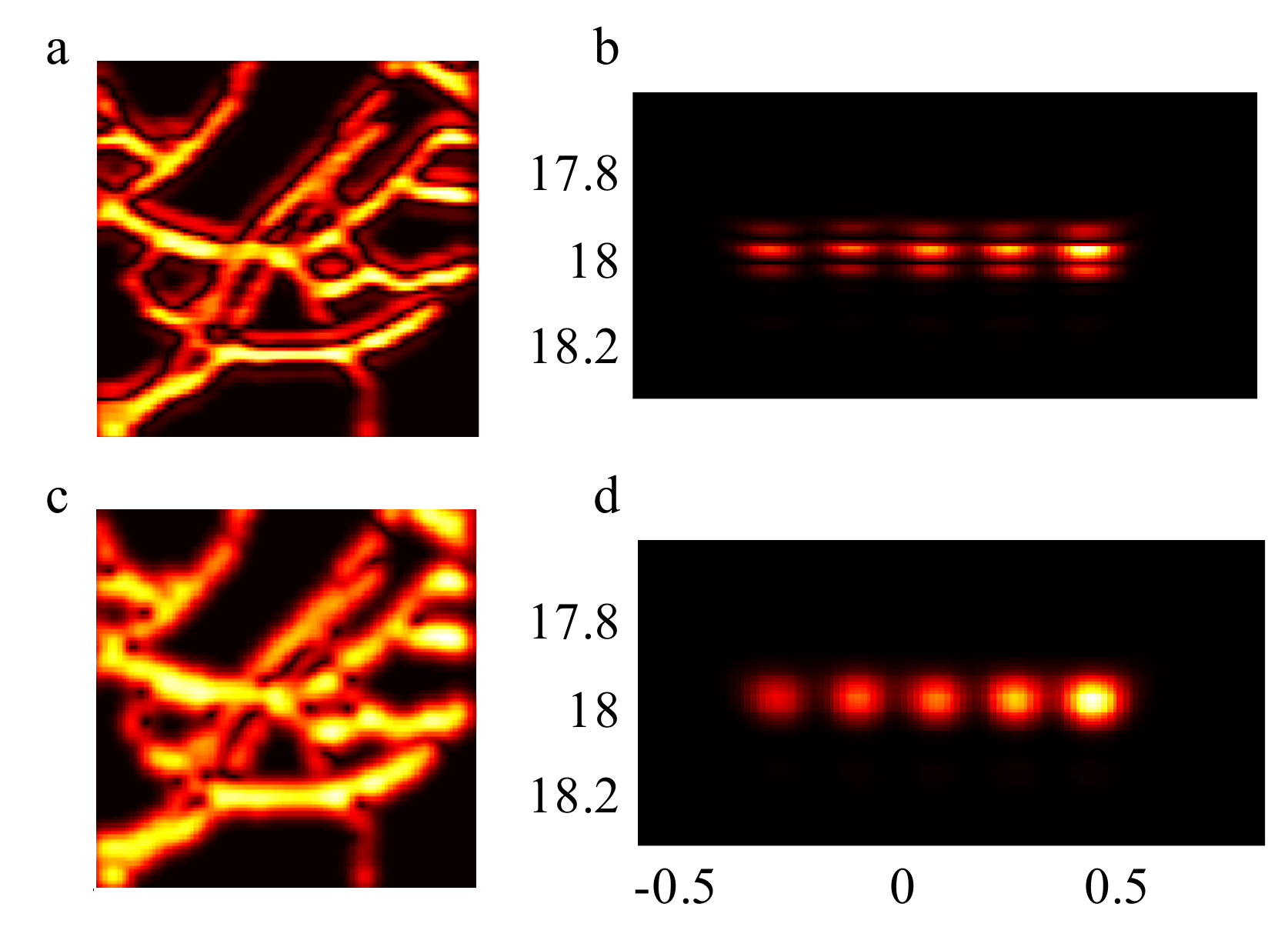}}
\caption{(a-b) $3^{rd}$-order cumulants, of real (a) numerical and (b) experimental data sets. (c-d) $3^{rd}$-order complex cumulants, for $p=1$ and $q=2$, of (c) numerical and (d) experimental data sets.}
\label{ResComplex}
\end{figure}

For both simulation (Fig. \ref{ResComplex}(a,c)) and experimental results (Fig. \ref{ResComplex}(b,d)), we compare the $3^{rd}$-order real cumulants Fig. \ref{ResComplex}(a-b) to the complex central ones Fig. \ref{ResComplex}(c-d), meaning $p=q$ if $n$ is even and $p=q\pm 1$ if $n$ is odd. 
By taking the magnitude of the complex cumulants, axial oscillations are discarded, as for conventional ultrasound imaging. Indeed, $K_{p,q}(\rr)$ is related to $ h(\mathbf{r})^{p}.h(\mathbf{r})^{*q}$ thanks to the additivity and homogeneity properties \cite{eriksson_essential_2010}. Then $|K_{p,q}(\rr)|$ is related to $|h(\mathbf{r})^{p}.h(\mathbf{r})^{*q}| = |h(\mathbf{r})|^{p+q}=|h(\mathbf{r})|^{n}$. Cumulants moduli with different $(p,q)$ values for a given $n^{th}$-order give different statistical estimates of the object to the $n^{th}$-power convoluted with the magnitude of the PSF to the $n^{th}$-power.
This analysis therefore provides artefact-free sub-acoustic resolution PA images.

\section{Conclusion}
We have demonstrated numerically and experimentally that flow-induced fluctuations in PA signals can be harnessed to provide resolution beyond the acoustic diffraction-limit. Our technique can be directly implemented on any conventional PA imaging system and with any illumination. The technique may be used to image blood micro-vessels that are otherwise blurred, by exploiting the natural flow of endogenous absorbers such as red blood cells.
Unlike previous approaches \cite{chaigne_super-resolution_2016,gateau_improving_2013}, the technique does not require coherent illumination, and is considerably less demanding in SNR requirement for fluctuations analysis, in cases where the absorbers dimensions are larger than the speckle grain size (i.e. the optical wavelength). In addition to the improved resolution, the technique increases the visibility of invisible structures in limited-view PA tomography \cite{gateau_improving_2013,dean2017dynamic}. Unlike localization based super-resolution techniques \cite{errico2015ultrafast}, in SOFI, the signal emitters distribution may be allowed to be dense, enabling a smaller required number of images and thus a reduction in acquisition time, which has to be tuned to reach a fair trade-off between spatial resolution, signal-to-noise ratio and time resolution \cite{dertinger_fast_2009}.

The basic statistical autocumulants analysis performed here can be potentially improved by calculating cross-cumulants and fluctuations at non-zero time-lags, which were shown to eliminate noise-originated correlations \cite{dertinger_fast_2009, bar2017fast}. Advanced image reconstruction algorithms \cite{hojman_photoacoustic_2017,murray_super-resolution_2017} can be employed to significantly improve image quality and spatio-temporal resolution, by taking into account knowledge of the imaging system response and any inherent structural sparsity of the target or correlation in the particle flow. In this respect, it is interesting to note that independent fluctuations occur between different blood vessels, but the fluctuations are spatio-temporally correlated between pixels of the same vessel, which can be used to smooth-out features along the vessel and to algorithmically improve reconstruction fidelity \cite{hojman_photoacoustic_2017}. 
We have performed the SOFI statistical fluctuations analysis on both real- and complex-valued beamformed images, and shown that the latter is an effective approach to discard the typical axial oscillations of the acoustic PSF. It may be possible to combine several cumulants of the same order for reducing even further imaging and statistical artefacts.
Finally, we note that the presented technique is based on statistical analysis of flowing absorbers, as such it may be combined with a Doppler based analysis \cite{fang2007photoacoustic,brunker2016acoustic} and related statistical analysis \cite{zhou2014calibration} to provide simultaneously information on flow velocity and particle concentration, in addition to the significantly improved resolution, from the same image dataset.\\

\textbf{Funding.}
This project has received funding from the European Research Council (ERC) under the European Union’s Horizon 2020 research and  innovation program (grants  no. 677909,  681514), and Human Frontiers Science Program. O.K. was supported by an Azrieli Faculty Fellowship. 

\textbf{Acknowledgment.} We thank Prof. Yonina Eldar and Oren Solomon for helpful discussions. We thank Mehdi Inglebert, Gwennou Coupier, Danielle Centani,  Philippe Marmottant and especially Sylvie Costrel for their help in the design and fabrication of the microfluidic samples.

\noindent 

\bibliography{Bib_Absorption_Fluctuation}

\end{document}